\documentclass[letterpaper, 10 pt, conference]{ieeeconf}  % Comment this line out if you need a4paper

\IEEEoverridecommandlockouts                              % This command is only needed if 
\title{\LARGE \bf
Unsupervised Motor Imagery Saliency Detection Based on Self-Attention Mechanism
}

\author{Navid Ayoobi$^{*}$ and Elnaz Banan Sadeghian$^{*}$% <-this % stops a space
\thanks{$^{*}$Navid Ayoobi and Elnaz Banan Sadeghian are with the Department of Electrical and Computer Engineering,
        Stevens Institute of Technology, Hoboken, USA.
        {(E-mail: \tt\small  nayoobi@stevens.edu,}
        {\tt\small ebsadegh@stevens.edu})
        }%
}
\usepackage{cite}
\usepackage{amsmath,amssymb,amsfonts}
\usepackage{algorithmic}
\usepackage{graphicx}
\usepackage{textcomp}
\usepackage{longtable}
\usepackage{multirow}
\usepackage{xcolor}
\usepackage{color, colortbl}
\definecolor{Gray}{gray}{0.92}
\definecolor{DarkGray}{gray}{0.8}
\usepackage{caption}
\usepackage{float}
\usepackage{diagbox}

\def\BibTeX{{\rm B\kern-.05em{\sc i\kern-.025em b}\kern-.08em
    T\kern-.1667em\lower.7ex\hbox{E}\kern-.125emX}}

\begin{document}
\maketitle
\thispagestyle{empty}
\pagestyle{empty}

%%%%%%%%%%%%%%%%%%%%%%%%%%%%%%%%%%%%%%%%%%%%%%%%%%%%%%%%%%%%%%%%%%%%%%%%%%%%%%%%
\begin{abstract}

Detecting the salient parts of motor-imagery electroencephalogram (MI-EEG) signals can enhance the performance of the brain-computer interface (BCI) system and reduce the computational burden required for processing lengthy MI-EEG signals. 
In this paper, we propose an unsupervised method based on the self-attention mechanism to detect the salient intervals of MI-EEG signals automatically.
Our suggested method can be used as a preprocessing step within any BCI algorithm to enhance its performance.
The effectiveness of the suggested method is evaluated on the most widely used BCI algorithm, the common spatial pattern (CSP) algorithm, using dataset 2a from BCI competition IV.
The results indicate that the proposed method can effectively prune MI-EEG signals and significantly enhance the performance of the CSP algorithm in terms of classification accuracy.

\end{abstract}

%%%%%%%%%%%%%%%%%%%%%%%%%%%%%%%%%%%%%%%%%%%%%%%%%%%%%%%%%%%%%%%%%%%%%%%%%%%%%%%%
\section{INTRODUCTION}

Brain-computer interface (BCI) bypasses the need for peripheral nervous and neuromuscular systems for communicating with surroundings and enables individuals to control external devices by directly translating neurophysiological brain signals into commands \cite{gaur2021sliding,sadeghian2007continuous}.
Among all signal acquisition methods in BCI, electroencephalography is employed in the vast majority of non-invasive systems.
Further, motor imagery electroencephalogram (MI-EEG) patterns captured when an individual imagines performing a movement are often used as the most discriminatory patterns for classification.
%Further, the neural activities occurring when an individual imagines performing a movement can be captured as motor imagery electroencephalogram (MI-EEG) signals. 
However, these MI-EEG signals have a poor signal-to-noise ratio (SNR).
Moreover, numerous forms of artifacts like electromyogram (EMG), power line interference, electrooculogram (EOG), eye movement, and blinks decline the SNR even further \cite{wei2019maximum}.
Due to the low SNR, unprocessed MI-EEG signals result in poor detection and classification performance.
Therefore, enhancing SNR is necessary for developing a reliable BCI system.

Artifact removal is one of the most popular preprocessing methods to increase the SNR of EEG signals.
In \cite{zou2019removing}, a method is proposed based on blind source separation to remove EMG artifacts from EEG data with a limited number of electrodes.
Chavez \textit{et al.} in \cite{chavez2018surrogate} adopt the wavelet decomposition and a resampling method to detect and remove ocular and muscular artifacts from a single-channel EEG across different scales.
However, these methods attempt to remove the artifacts from the entire EEG signal where the user performance can vary while performing an MI task varies during a trial.
Therefore, using the entire trial is not optimally suitable for classification and degrades the classification accuracy.
Moreover, processing a lengthy EEG is computationally inefficient.
Numerous efforts have been conducted to address this issue by using a time window to detect the most salient interval in a trial for MI classification \cite{gaur2021sliding,hsu2007wavelet}:
Gaur \textit{et al.} in \cite{gaur2021sliding} use a $2$-second sliding window and split the EEG signal into nine intervals where the CSP algorithm is used to extract features and nine linear discriminant analysis (LDA) classifiers are trained separately for each individual interval. 
The longest consecutive repetition is then used to choose the best label among the nine labels generated by different LDAs.
Hsu \textit{et al.} in \cite{hsu2007wavelet} propose a method based on continuous wavelet transform and Student’s two-sample t-statistics to detect one salient interval within every $4$-second trial using a $1$-second searching window. 
However, these approaches do not detect the entirety of salient intervals which can be several and of varying lengths. 
%However, using a single fixed window cannot detect the entirety of salient intervals that can be several and with varying lengths.
Furthermore, they require labeled data, and their effectiveness depends on the selected classifier's ability to separate them.

This paper proposes a method based on the self-attention mechanism \cite{bahdanau2014neural} to automatically extract several salient intervals in every MI-EEG trial. Further, the method is unsupervised and therefore it does not require any labeled data.
The proposed method can be applied to MI-EEG signals as a preprocessing step within any BCI algorithm to improve the SNR and enhance its performance.
The proposed architecture is comprised of an input embedding layer, an encoder-decoder network, a self-attention layer, and a dense layer.
We utilize the encoder-decoder structure to reconstruct the input, yielding an unsupervised architecture.
The self-attention layer assigns a weight to each input time sample and aids the decoder to precisely reconstruct the input based on those time samples with the highest weights representing the salient parts of each trial.
Consequently, the proposed method can detect different numbers of salient intervals with varying lengths.
%The encoder converts each embedded input into a vector.
%The decoder and the dense layer reconstruct the input from the encoded vector using the contexts generated by the self-attention layer.
%This layer learns to assign higher weights to the salient intervals aiding the decoder to precisely reconstruct the input.
After the training phase, the decoder part is removed from the network.
The trained encoder and self-attention layer is then used to extract salient intervals of unseen testing trials.

To evaluate our approach, we use dataset 2a from BCI competition IV to extract the salient intervals of the MI trials. 
We compare the performance of the most widely used algorithm in BCI, the common spatial pattern (CSP) \cite{blankertz2007optimizing} algorithm, with and without our proposed method.
The results show that the proposed method can effectively extract the salient intervals of MI-EEG signals and significantly improve the CSP performance in terms of classification accuracy.

The remainder of this article is organized as follows.
Section \ref{dataset} introduces the dataset.
Section \ref{method} elaborates the proposed method and architecture.
Section \ref{results} presents the simulation results and discussion.
Section \ref{conclusion} concludes this article.

\section{Dataset} \label{dataset}

\begin{figure}[t]
    \centering
    \includegraphics[width=\columnwidth]{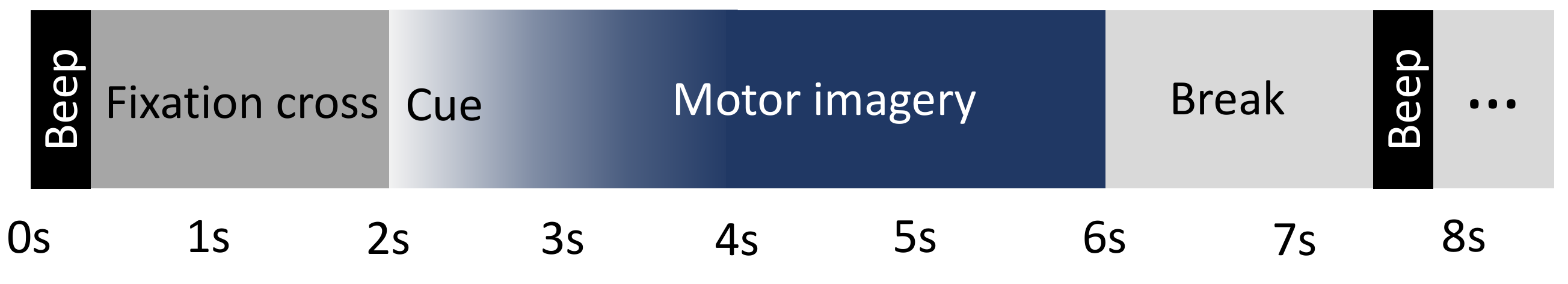}
    \caption{\small Timing scheme of each trial in dataset 2a from BCI competition IV.}
    \label{fig:trial}
\end{figure}

To evaluate our proposed method, we use dataset 2a from BCI competition IV \cite{brunner2008bci} that contains the EEG data of nine healthy subjects.
This dataset consists of movement imaginations of left and right hands, feet, and tongue.
The EEG signals are recorded using twenty-two Ag/AgCl electrodes and sampled at $250$ Hz.
The timing scheme of each trial is depicted in Fig. \ref{fig:trial}.
In this paper, we only utilize the left-hand and right-hand trials. 
We also extract all our trials from the second $2$ after the cue to the second $6$.

\section{Methods} \label{method}

\begin{figure}[t]
    \centering
    \includegraphics[width=\columnwidth]{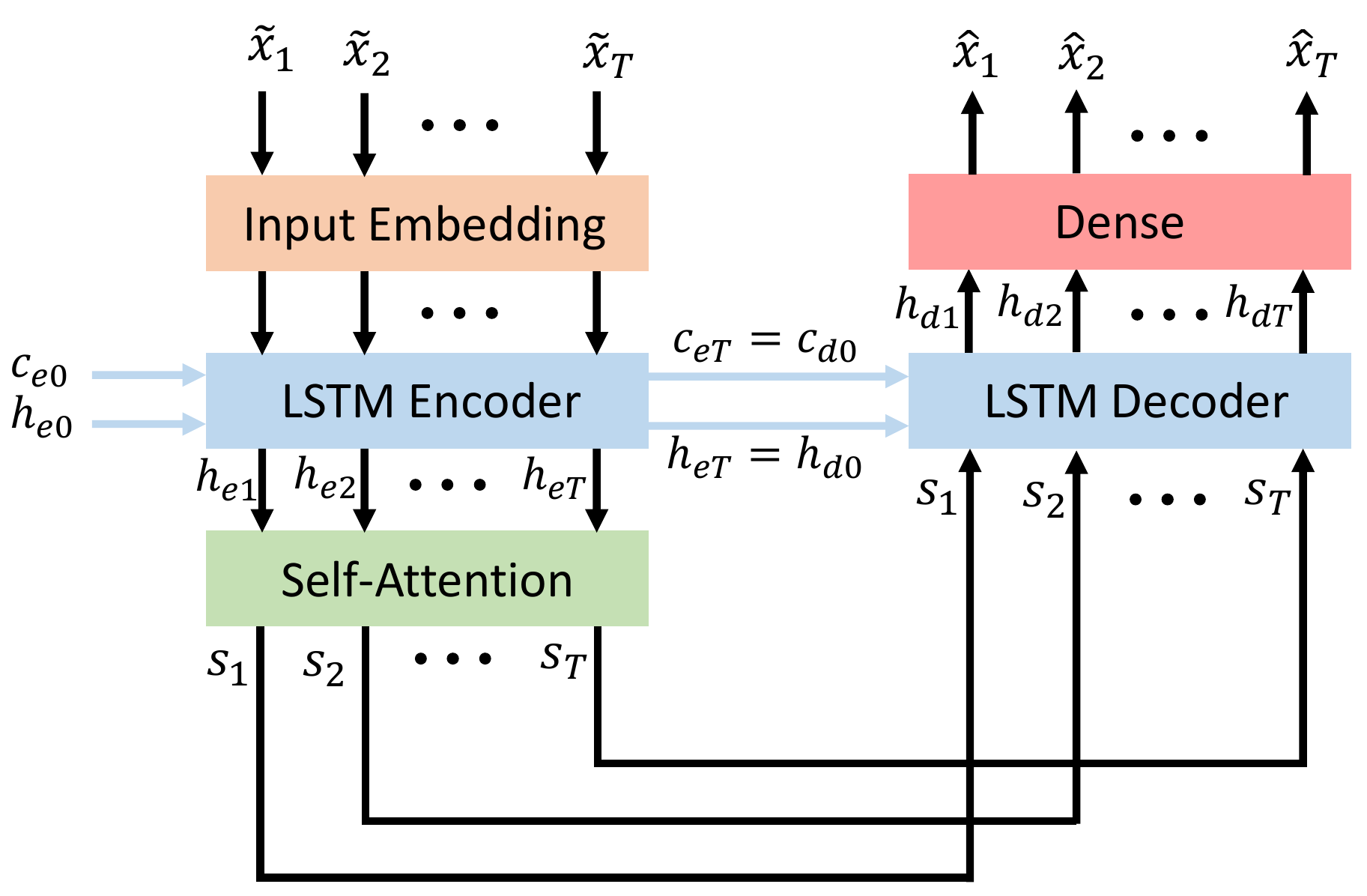}
    \caption{\small The block diagram of the proposed method.}
    \label{fig:net}
\end{figure}

In this section, we elaborate on our proposed unsupervised method for detecting the salient parts of EEG trials to improve classification accuracy.
The suggested architecture consists of four main parts: input embedding, long short-term memory (LSTM) encoder-decoder, self-attention, and dense layer.
The encoder-decoder structure helps the self-attention layer learn the salient parts of the EEG trials in an unsupervised method by reconstructing the input.
The proposed architecture is shown in Fig. \ref{fig:net}.

\subsection{The proposed architecture}

\textbf{Input embedding:} 
Given a multivariate signal as input, the embedding layer captures the dependencies between different channels \cite{song2018attend} in EEG signals and creates a representation vector for each time sample.
We employ a 1D convolutional layer with $m$ numbers of $d$-dimensional kernels.
This layer produces an $m$-dimensional vector for each time sample by looking at current and $d{-}1$ next time samples. 
We utilize zero padding to ensure that the input and output sequences have the same length.

\textbf{LSTM encoder:} 
LSTM as a variant of recurrent neural networks can capture long temporal dependencies in sequential data \cite{greff2016lstm}.
We employ LSTM units to build our encoder and decoder.
The encoder processes the input sequence of length $T$ and summarizes the temporal information of each time sample in two $h$-dimensional vectors, the hidden state $h_{e}$ and the cell state $c_{e}$.
The produced hidden states are then fed to a self-attention layer.
After $T$ recurrences, the whole sequence is summarized in final encoder hidden state $h_{eT}$ and cell state $c_{eT}$.
These two vectors are fed to the decoder's initial hidden state $h_{d0}$ and cell state $c_{d0}$.

\textbf{Self-attention:} 
Attention is originally proposed in \cite{bahdanau2014neural} to improve the performance of an encoder-decoder in a machine translation.
It alleviates the bottleneck problem of using the fixed-length vector from which the decoder produces the output sentence by allowing the model to automatically seek source sentence parts that are significant for predicting a target word.
Self-attention, also known as intra-attention, is an attention mechanism that links distinct points in a single sequence to calculate a representation of the sequence \cite{vaswani2017attention}.
In this work, we adopt the self-attention mechanism to detect the salient parts of each MI-EEG trial.
In our self-attention layer, the $i$-th encoder hidden state $h_{ei}{=} \{h_{ei,1}, ...\,,h_{ei,h}\}$ is mapped to an $n_k$-dimensional query vector $q_i$, an $n_k$-dimensional key vector $k_i$, and an $n_v$-dimensional value vector $v_i$ by multiplying $h_{ei}$ with three matrices $W_q$, $W_k$ and $W_v$, respectively:
\begin{equation}
    q_i = W_q h_{ei}
\end{equation}
\begin{equation}
    k_i = W_k h_{ei}
\end{equation}
\begin{equation}
    v_i = W_v h_{ei}
\end{equation}

These vectors are used to calculate a context vector in each position in the sequential data.
Here, dot-product attention \cite{vaswani2017attention} is used to compute the context vectors.
The context vector at $i$-th position is a weighted sum of value vectors in all positions
\begin{equation}
    c_i = \sum_{t=1}^{T}\alpha_{i,t} v_t ,
\end{equation}

where the attention weight $\alpha_{i,t}$ measures how well the $i$-th encoder hidden state aligns with the encoder hidden state at position $t$.
The attention weights are computed as the inner product of the query with the corresponding key.
After computing the weights associated with all keys, a softmax function is used to make the weights sum up to one.
If all query vectors, key vectors, and value vectors are packed together into $\mathbf{Q}$, $\mathbf{K}$, and $\mathbf{V}$ matrices, the context matrix $\mathbf{C}$ is calculated as follows
\begin{equation}
    \mathbf{C} = softmax(\mathbf{Q}\mathbf{K}^T)\mathbf{V} .
\end{equation}

\textbf{LSTM decoder and dense layer:}
The LSTM decoder accepts the final hidden state and cell state of the encoder as its initial hidden state and cell state.
The context vectors generated by the self-attention layer are fed to the decoder inputs.
The decoder outputs are then passed through a dense layer to increase the dimensionality of decoder outputs to the number of EEG channels.
The decoder and dense layer learn to reconstruct the original MI trial by compelling the self-attention layer to pay more attention to certain parts of the MI trial that are more significant for reconstructing the input.

\subsection{Training and Testing stages}

\textbf{Training stage:}
To improve the performance of the proposed network in reconstructing the input EEG, $x$, we randomly select $p_1$ time samples of $x$ and set $p_2$ feature values to zero to obtain $\Tilde{x}$.
In the training stage, $\Tilde{x}$ is used as the training trial, and its corresponding original EEG signal $x$ is used as the ground truth.
The mean squared error is used as the loss function
\begin{equation}
Q =\frac{1}{Tn_c} \sum_{t=1}^T\sum_{i=1}^{n_c} ( \hat x_{t}^{(i)} - x_t^{(i)})^2,
\end{equation}
where $n_c$, $\hat x$ and $x$ are the number of EEG channels, the reconstructed EEG, and the original EEG, respectively.

\textbf{Testing stage:}
The decoder and dense layer are removed from the network for the testing stage.
A test trial is passed through the embedding layer, the encoder, and the self-attention layer.
The computed attention weights $\{\alpha\}$ are used to detect the salient parts of the test trial.
Assume that the attention weights of a test trial are represented in a matrix $ \Lambda_{T \times T}$. The attention at position $t$, $a_t$, is computed as follows
\begin{equation}
    a_t = \frac{1}{T} \sum_{j=1}^{T} \Lambda_{j,t}.
\end{equation}

The computed attentions form an attention vector $A=[a_1,\,... \,,a_T]$.
We segment the attention vector into $n$ intervals for separating salient intervals.
$r$ out of $n$ intervals with the highest average attention are selected.
These two hyperparameters are tuned in the cross validation procedure.
The time samples corresponding to the selected attention intervals are then concatenated to form the pruned trial.

\section{Results and discussion}\label{results}
\begin{figure*}[t]
    \centering
    \includegraphics[width=1.8\columnwidth]{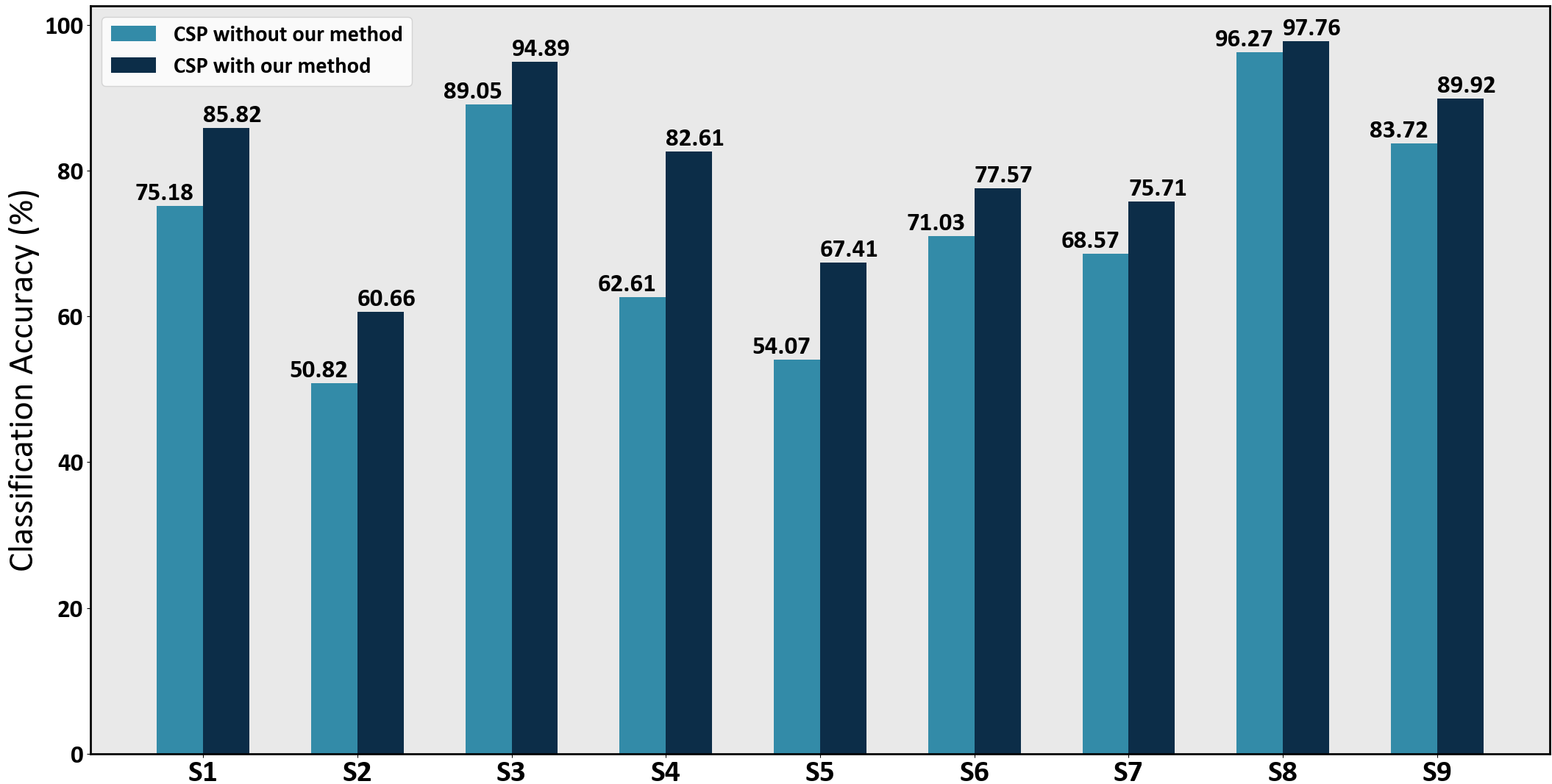}
    \caption{\small Comparison of the performance of the CSP algorithm for different subjects in the dataset before and after applying our proposed method in terms of classification accuracy.}
    \label{fig:acc}
\end{figure*}

In all our experiments, the MI-EEG signals are bandpass filtered between $4$ Hz and $40$ Hz with a fifth-order Butterworth filter.
For all subjects, the number of kernels $m$, kernel dimension $d$, LSTM hidden state dimension $h$, key-query dimension $n_k$, the dimension of the value vector $n_v$, parameter $p_1$, and parameter $p_2$ are set to $5$, $4$, $4$, $4$, $4$, $0.6$, and $0.4$, respectively.
A deep neural network with three hidden layers (with 5,10, and 15 nodes) is used for the dense layer.  
The parameters $r$ and $n$ are tuned by 5-fold cross validation for each subject.
In addition, we use three pairs of spatial filters in the CSP algorithm \cite{blankertz2007optimizing}.

\subsection{The effect of the proposed method on the classification performance}

To evaluate the proposed method, we compare the performance of the CSP algorithm under two scenarios: 1) with and 2) without applying our method to the MI-EEG trials.

In the first scenario, we train our proposed architecture with the $4$-second training trials.
After the training stage, we use the encoder and self-attention layer to extract the salient intervals from the training and testing trials.
The spatial filters of the CSP algorithm are then designed using the pruned training trials.
The accuracy of the CSP algorithm is computed using the pruned testing trials.
In the second scenario, the $4$-second training trials and testing trials are directly used to design the spatial filters and obtain the accuracy of the CSP algorithm, respectively.
In both scenarios, linear discriminant analysis is used as the classifier.

The results are shown in Fig. \ref{fig:acc} for all the subjects.
We observe that the classification accuracy is improved approximately by $14.2\%$, $19.4\%$, $6.6\%$, $31.9\%$, $24.7\%$, $9.2\%$, $10.4\%$, $1.5\%$, and  $7.4\%$ for subject one through nine, respectively.
On average, the proposed method enhances the classification accuracy of the CSP by about $13.9\%$ across all subjects.

In addition, the ratio of the length of the pruned trials to the length of the unpruned trials is $0.7$, $0.65$, $0.45$, $0.7$, $0.55$, $0.7$, $0.75$, $0.75$, and $0.75$ for subject 1 through 9, respectively.
Hence, the proposed method reduces the computational burden required for processing EEG signals and increases the classification accuracy simultaneously.

\subsection{The effect of segment length on attention vector splitting}

\begin{figure}[t]
    \centering
    \includegraphics[width=\columnwidth]{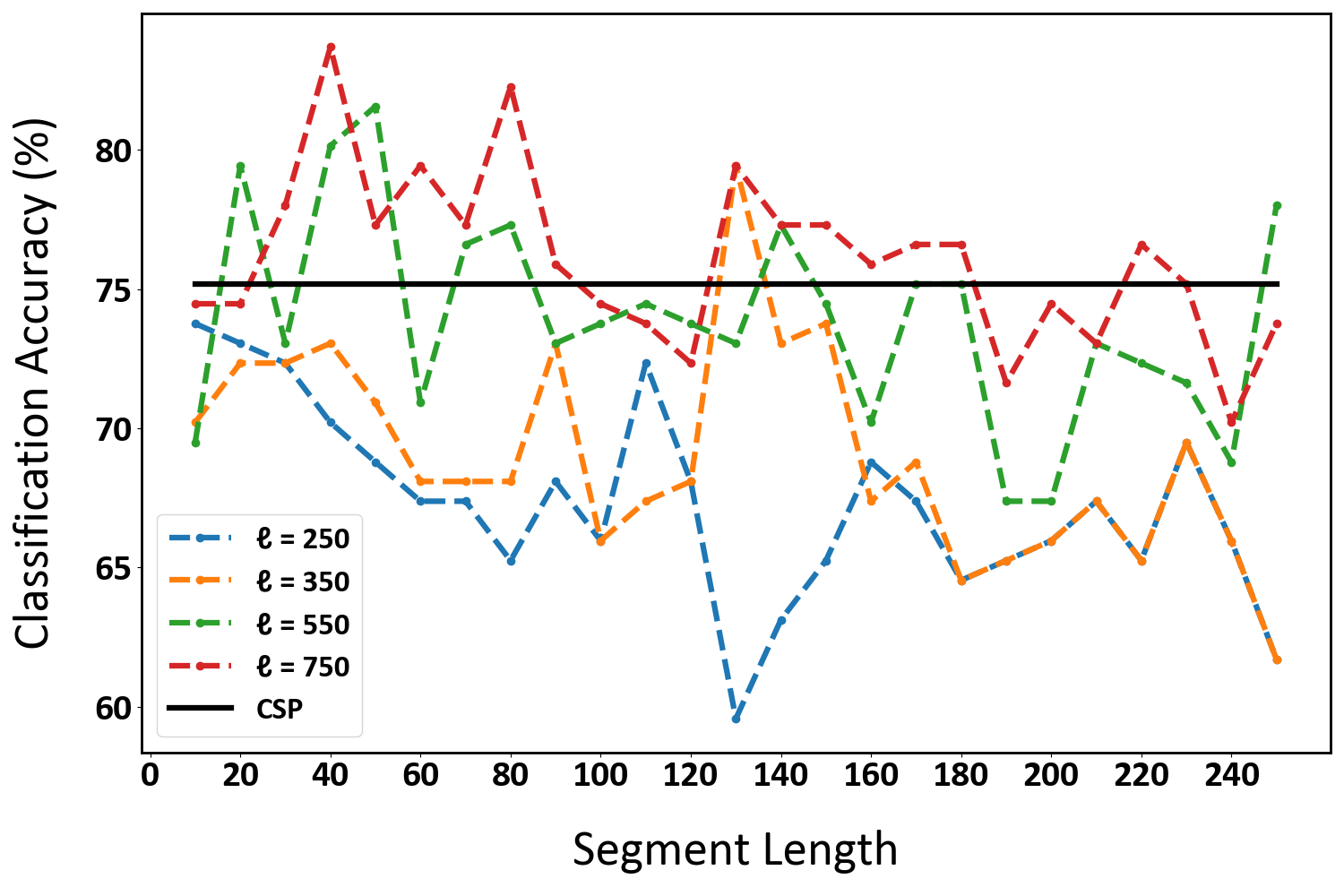}
    \caption{\small The effect of different segment lengths on attention vector splitting for subject one.}
    \label{fig:subject1}
\end{figure}

This section evaluates the effect of different segment lengths $T/n$ used for splitting the attention vector to extract the salient intervals.
Fig. \ref{fig:subject1} compares the results in terms of classification accuracy for subject one in the dataset versus the number of time samples in each interval, $T/n$.
$r$ intervals with the highest average attention are concatenated to form a trial containing $\ell{=}rT/n{=}750$ (red dashed line), $\ell{=}550$ (green dashed line), $\ell{=}350$ (orange dashed line), and $\ell{=}250$ (blue dashed line) time samples.
The black horizontal line shows the classification accuracy of the CSP algorithm without our method.
We observe that restricting the concatenated signal to have a very short length ($\ell{=}350$ and $\ell{=}250$) results in lower accuracy compared to the CSP method alone.
As shown the best curves correspond to $\ell{=}750$ and $\ell{=}550$. These curves prove that using the entire lengths of the trials  does not always achieve the best obtainable accuracy.

In addition, the results suggest that shorter segment lengths $T/n$ can generally provide higher accuracy than longer segment lengths.
In other words, using multiple time windows instead of a single time window \cite{gaur2021sliding,hsu2007wavelet} results in detecting more salient intervals in a trial.
The proposed method may be employed as a prepossessing step in MI-BCI studies for enhancing motor imagery classification.

\section{Conclusion}\label{conclusion}

In this paper, we present an unsupervised method based on the self-attention mechanism to automatically detect the salient intervals of MI-EEG trials.
We test our proposed method using the CSP algorithm on the dataset 2a from BCI competition IV.
The results show that pruning the MI-EEG trials by the proposed method can effectively improve the SNR and reduce the computational burden in BCI systems.
The results also show that the average classification accuracy of the CSP algorithm is improved by approximately $13.9\%$ across all subjects. 
The suggested approach may be used as a preprocessing step before any BCI algorithm to improve its performance.

\bibliographystyle{IEEEtran}
\bibliography{./root.bib}

\end{document}